\documentclass[letter]{aa}
\usepackage{amsmath}  
\usepackage{gensymb}
\usepackage{xcolor}
\usepackage{graphicx}
\usepackage{txfonts}
\usepackage{lipsum}
\usepackage[colorlinks=true,allcolors=blue]{hyperref}
\usepackage{siunitx}
\usepackage{subcaption}     
\usepackage{etoolbox}
\pretocmd{\appendix}{\crefalias{section}{appendix}}{}{}
                    
\usepackage{lscape}       
                         
\usepackage{placeins}

\usepackage{cleveref}

\usepackage{multirow}
\newcommand{\kms}{\,km\,s$^{-1}$}

\begin{document}

   \title{HCN detection in the Kreutz sungrazing comet C/2026 A1 (MAPS) shortly before its disruption\thanks{Based on observations carried out with the IRAM 30-m telescope. IRAM is supported by INSU/CNRS (France), MPG (Germany) and IGN (Spain).}}

   \subtitle{}

\titlerunning{C/2026 A1 (MAPS) shortly before disruption}
\authorrunning{A. Godard Palluet et al.}

   \author{Amélie Godard Palluet\inst{1} \fnmsep \thanks{Corresponding author: {\ttfamily amelie.godard@cab.inta-csic.es}},
	Dominique Bockelée-Morvan\inst{2},
	Nicolas Biver\inst{2},
	Raphael Moreno\inst{2},
    Martin Cordiner\inst{3,4},
	David Haasler\inst{1,5},
	Jacques Crovisier\inst{2},
	\and Víctor M. Rivilla\inst{1}
}

\institute{Centro de Astrobiología (CAB), CSIC-INTA, Carretera de Ajalvir km 4, Torrejón de Ardoz, 28850, Madrid, Spain
	\and LIRA, Observatoire de Paris, PSL Research University, CNRS, Sorbonne Université, Université de Paris, 5 place Jules Janssen,
	F-92195 Meudon, France
    \and Astrochemistry Laboratory, NASA Goddard Space Flight Center, 8800 Greenbelt Road, Greenbelt, MD 20771, USA
    \and Department of Physics, Catholic University of America, Washington, DC 20064, USA
	\and Departamento de Física de la Tierra y Astrofísica, Facultad de Ciencias Físicas, Universidad Complutense de Madrid,
	28040 Madrid, Spain}

  \abstract
{Kreutz sungrazing comets are characterised by extremely small perihelion distances ($q < 3.45 \, R_\odot$). These objects are of particular interest because their intense activity and fragmentation can expose material from deeper within the nucleus, providing unique access to their bulk chemical composition rather than merely their surface layers.}
{The discovery of C/2026 A1 (MAPS) three months before its perihelion offered a rare opportunity to investigate the composition and activity of a Kreutz sungrazing comet. We present the first millimetre wavelength study of this type of comet.}
{We observed C/2026 A1 (MAPS) with the IRAM 30-m telescope from April 1 to April 3 2026 at heliocentric distances $r_{h}$ = 0.10--0.23\,au, the three days preceding its perihelion and full disintegration. We analysed the molecular emissions using non-local thermodynamic equilibrium radiative transfer models to derive molecular production rates.}
{We report the first HCN detection in a Kreutz sungrazer through its (3--2) transition at 265.886\,GHz, enabling us to characterise the variability of its activity. We witnessed two outbursts, on April 1 and April 3. The second one, occurring the day before the disintegration of C/2026 A1 (MAPS), led to the release of 0.4\% to 1.2\% of the total cometary mass, thus suggesting a pre-disruption before entering the Roche limit region. We also obtained upper limits for HNC, CS, H$_2$CO, CO, OCS and CH$_3$OH.}
{}


   \maketitle
   \nolinenumbers

\section{Introduction}

Kreutz sungrazers are a dynamical population of comets with extremely close solar passages (within Roche limit, i.e. heliocentric distance $r_{h}<0.016$\,au). All Kreutz sungrazers are thought to be fragments of a single Oort cloud comet that would have broken up during a perihelion passage at very small $r_{h}$, or throughout the orbit \citep{Jones2018}. In the 21st century, only three Kreutz sungrazers have been detected from the ground: C/2011 W3 (Lovejoy, the Great Comet of 2011), C/2024 S1 (ATLAS), and C/2026 A1 (MAPS). 

The Kreutz sungrazing comet C/2026 A1 (MAPS) (hereafter, C/2026 A1) was discovered in January 2026 by A. Maury, G. Attard, D. Parrott and F. Signoret as part of the MAPS survey \citep{Maury2026A1}. It was detected at the largest heliocentric distance pre-perihelion yet recorded for a Kreutz sungrazing comet ($r_{h}$ = 2.1\,au). With an orbital period of $\sim$1700\,years, it was expected to reach an exceptionally small perihelion distance of 0.006\,au on April 4.60 2026 coordinated universal time (UTC). Soon after its discovery, observations were conducted with the James Webb Space Telescope (JWST) at $r_{h}$ = 1.6\,au, providing the first direct constraint on the nucleus of such an object, with an estimated diameter of 0.4\,km \citep{Zhang2026}. 

On April 4, $\sim$2\,hours before perihelion, C/2026 A1 fully disintegrated. This event was captured by the Solar and Heliospheric Observatory (SOHO) spacecraft using the Large Angle and Spectrometric Coronagraph (LASCO). The comet neared the Sun, disappeared behind the coronagraph disk, and shortly afterwards a cloud of cometary dust appeared on the other side of the disk, the remnant of its final disintegration. The Solar Terrestrial Relations Observatory (STEREO) SECCHI-Cor2-A instrument observed the same event from a different angle, directly showing the comet turning to dust as it approached the Sun.\footnote{\url{https://science.nasa.gov/blogs/the-sun-spot/2026/04/16/nasa-heliophysics-spacecraft-witness-comets-demise/}}

    \begin{figure*}[t!]
    \centering
     \renewcommand{\arraystretch}{1.1}
    \begin{minipage}{\linewidth}
    \captionof{table}{Integrated line intensities ($\int T_{\rm mb}{\rm d}v$) of the HCN(3--2) transition (at 265.886\,GHz) for April 1--3 2026, with the corresponding gas expansion velocity ($v_{\rm exp}$), molecular production rates ($Q_{\rm HCN}$), heliocentric distance ($r_h$), and total visual magnitude ($m_1$).}
    \label{table:line_int_hcn}
    \centering
    \begin{tabular}{r c c c c c c c} 
    \hline\hline  
    UTC date      &  $r_{h}$ & $m_{1}$$^{(a)}$ & $\int T_{\rm mb}{\rm d}v$  &  S/N    & $v_{\rm exp}$   & Doppler shift  & $Q_{\rm HCN}$ \\  
    (yyyy/mm/dd.dd) &  (au)  &  & (mK\,km\,s$^{-1}$) &         & (km\,s$^{-1}$)    & (km\,s$^{-1}$) &( $10^{25}$ mol.\,s$^{-1}$) \\  
    \hline 
     2026/04/01.41 - 01.58$^{(b)}$ & 0.233--0.224 & 9.4   & 130 $\pm$ 17 & 7.7 & 0.85 $\pm$ 0.14 &  0.11 $\pm$ 0.11 & 6.9 $\pm$ 0.9\\
    \cline{2-8}
     01.41 - 01.45      & 0.233--0.231  &  9.5 &  166 $\pm$ 39 &  4.3 & 1.21 $\pm$ 0.24  &  $-0.14$ $\pm$ 0.21  & 8.3 $\pm$ 1.7  \\ 
     01.45 - 01.47      & 0.231--0.230  &  9.4 &  218 $\pm$ 52 &  4.2 & 0.60  $\pm$  0.29  &   0.046 $\pm$ 0.16  & 8.0 $\pm$ 1.8  \\ 
     01.47 - 01.58$^{(c)}$ & 0.230--0.224  &  9.4 &  96 $\pm$ 22  & 4.4 &  0.84 $\pm$ 0.24  &   0.13 $\pm$ 0.18  & 5.9 $\pm$ 1.1 \\ 
     \hline
     2026/04/02.49 - 02.70 &  0.176--0.164  & 8.2 &   93 $\pm$ 18 & 5.3  & 0.80 $\pm$ 0.23 & 0.56 $\pm$ 0.16 & 7.6 $\pm$ 1.4 \\
     \hline
     2026/04/03.45 - 03.56 &  0.116--0.109  & 6.2 &  218 $\pm$ 23 & 9.8 & 1.17 $\pm$ 0.14 & 0.027 $\pm$ 0.11 & 36 $\pm$ 4\\ 
    \cline{2-8}
     03.45 - 03.49 &  0.116--0.113  & 6.3 &  148 $\pm$ 34 &  4.3 & 1.10  $\pm$ 0.30 & 0.20 $\pm$ 0.22 & 25 $\pm$ 6 \\ 
     03.49 - 03.50 &  0.113--0.111  & 6.2 &  173 $\pm$ 42 &  4.1 & 1.00  $\pm$ 0.29 & $-0.06$ $\pm$ 0.22 & 34 $\pm$ 7 \\     
     03.51 - 03.53 &  0.111--0.110  & 6.2 &  355 $\pm$ 57 &  6.2 & 0.98  $\pm$ 0.20 & $-0.06$ $\pm$ 0.15 & 59 $\pm$ 9  \\   
     03.53 - 03.56 &  0.110--0.109  & 6.1 &  290 $\pm$ 62 &  4.7 & 1.52  $\pm$ 0.28  & 0.12 $\pm$ 0.22 & 45 $\pm$ 8  \\  
    \hline
    \hline
    \end{tabular}
    \begin{center}
    \end{center}\vspace{-0.7cm}
    {\small
    \begin{flushleft}
    $^{(a)}$ Values were taken from the Minor Planet Center: \url{https://cgi.minorplanetcenter.net}.\\
    $^{(b)}$ These scans have an average pointing offset of 1.9\arcsec as some of them were acquired when performing a coarse map aimed at estimating the position of maximum HCN emission. The offset was accounted for in the beam convolution used to derive molecular production rates.\\
    $^{(c)}$ These scans were captured with an average offset of 3.2\arcsec.
    \end{flushleft}
    }
    \end{minipage}

    \vspace{0.2cm}

        \includegraphics[width=\linewidth]{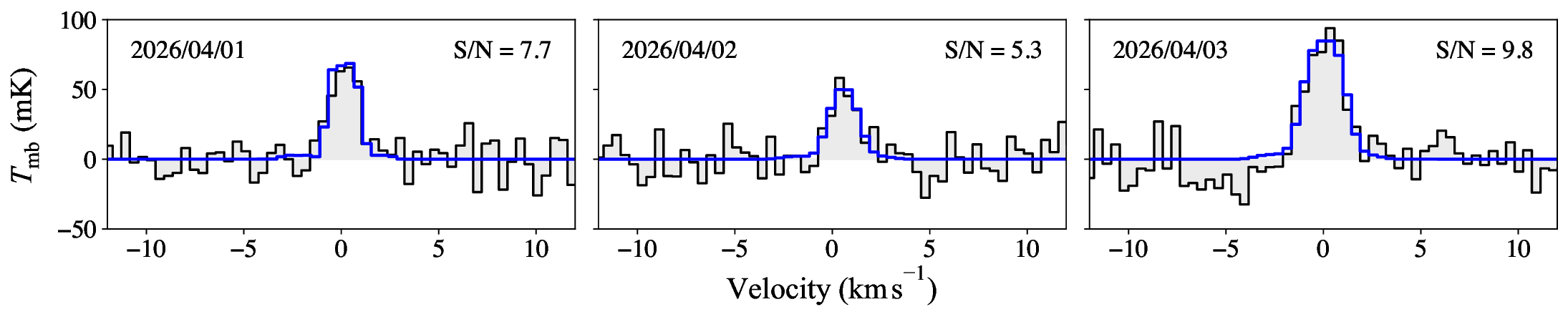} 
    \vspace{-6mm} 
    \captionof{figure}{Observed (grey lines) and modelled (blue lines) main beam temperature ($T_{\rm mb}$) in the comet velocity rest-frame for the HCN(3--2) emission line on April 1 (left panel), April 2 (central panel) and April 3 (right panel). The UTC date (yyyy/mm/dd.dd, top left corner) and velocity-integrated signal-to-noise ratios (S/N, top right corner) are indicated in each panels.}
    \label{fig:hcn_detection_daily}
    \end{figure*}

 We observed C/2026 A1 in the three days immediately preceding its disintegration, from April 1 to April 3 ($r_{h}$ = 0.11--0.23\,au), using the Institut de radioastronomie millimétrique (IRAM) 30-m telescope. This study complements previous observations at (sub)millimetre wavelengths of highly active comets at very small heliocentric distances, including C/2002 V1, and C/2006 P1, whose perihelion distances were 0.10\,au and 0.17\,au and that were observed at 0.11--0.13\,au and 0.21--0.25\,au, respectively \citep{Biver2011}.

In this letter, we report the detection of the HCN(3–2) line at 265.886\,GHz, the first millimetre molecular detection in a Kreutz sungrazer. We monitored the comet activity through the HCN(3--2) line on the three days and witnessed two outbursts that may have contributed to its final disintegration.


\section{Methods}
\subsection{Observations with IRAM 30-m}

    The observations were conducted with the IRAM 30-m telescope (Sierra Nevada, Spain) over three consecutive days, from April 1 to 3, 2026 under Director Discretionary Time (DDT E03-25). During these observations, comet C/2026 A1 was near its closest approach to the Sun, with $r_{h}$ ranging from 0.23\,au (April 1) to 0.10\,au (April 3), and its geocentric distance ranging from $\Delta$ = 1.10\,au to $\Delta$ = 1.04\,au.

    The observations were conducted in wobbler switching mode, with two Eight MIxer Receiver (EMIR) setups, both in the E230/1mm band, connected to the Fast Fourier Transform Spectrometer (FTS, resolution of 200\,kHz), and VESPA autocorrelator (resolution of 40\,kHz) in parallel. One setup was optimized for HCN(3--2) at 265.886\,GHz in the upper sideband (USB), and CH$_3$OH lines between 250\,GHz and 254\,GHz in the lower sideband (LSB). The second setup was designed to target CS(5--4) at 244.936\,GHz and CH$_3$OH lines near 241\,GHz in the USB, as well as CO(2--1) at 230.538\,GHz, and H$_2$CO($3_{12}$--$2_{11}$) at 225.697\,GHz in the LSB. The observing sequence included a series of $\sim$6\,min scans. Pointing and focus corrections were performed on Venus every hour. Additional informations can be found in \Cref{sec:obs_conditions}.

   \subsection{Non-LTE radiative transfer modelling}\label{sec:non-LTE}

    \begin{figure*}[t]
        \includegraphics[width=\linewidth]{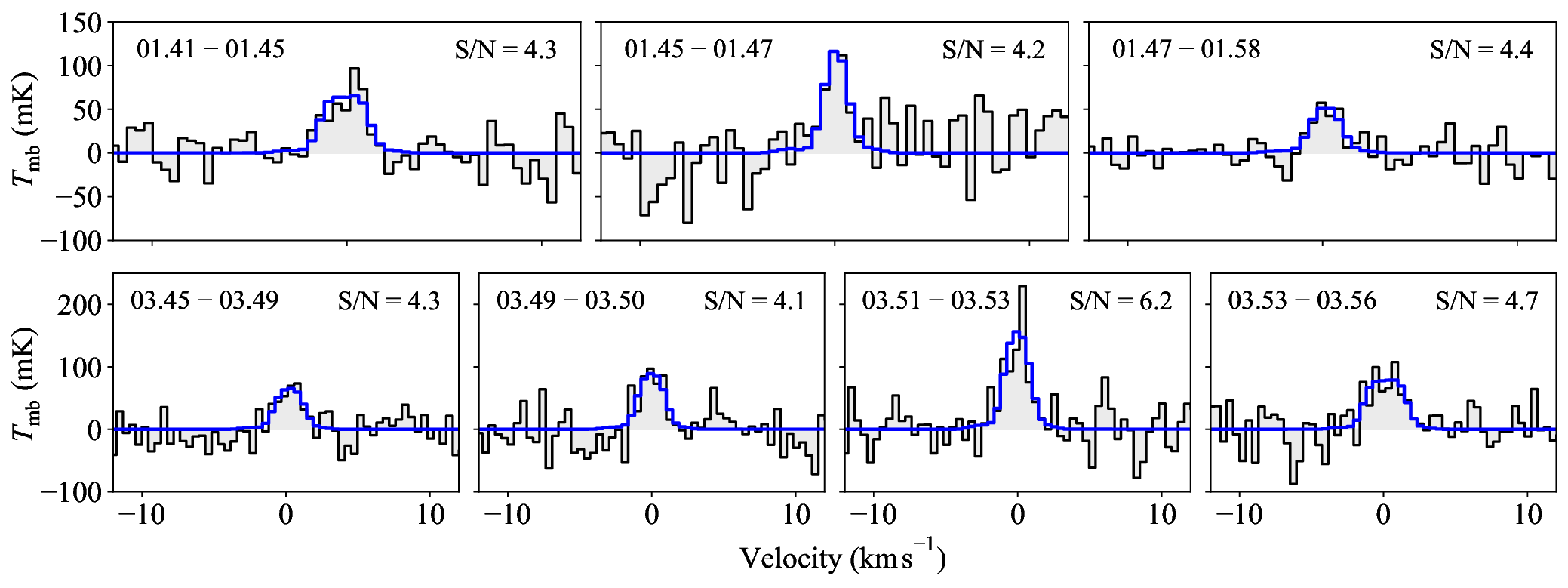} 
        \vspace{-6mm} 
    \captionof{figure}{Time evolution of the HCN(3--2) spectra observed on April 1 (top panels) and April 3 (bottom panels), see caption of \Cref{fig:hcn_detection_daily}. The UTC time (dd.dd, top left corner) and velocity-integrated signal-to-noise ratios (S/N, top right corner) are indicated in each panel.}
    \label{fig:hcn_detection_detail}
    \end{figure*}

    The gas within the coma usually spans a broad density range (10$^{1}$--10$^{12}$\,cm$^{-3}$) due to expansion from the nucleus \citep{Despois2006}. The probed region is thus a mix of gas under local thermodynamic equilibrium (LTE) and non-LTE conditions. Moreover, photodissociation by the solar radiation field becomes very efficient at small heliocentric distances. HCN survives $\sim$3100\,km from the nucleus at $r_{h}=0.2$\,au, and $\sim$600\,km at $r_{h}=0.1$\,au for an expansion velocity of the gas $v_{\rm exp}=1$\,km\,s$^{-1}$. For a water production rate $Q_{\rm H_{2}O} = 1 \times 10^{29}$\,mol.\,s$^{-1}$, the LTE region breaks down at $\sim$1000\,km from the nucleus. A non-LTE approach is thus expected to better encompass the physics across all regions of the coma.
        
    For the analysis of the emission lines, we employ the 1-dimensional version of the non-LTE radiative transfer model SUBLIME (SUBlimating gases in LIME, \citealt{Cordiner2022}). It is a modified version of the LIne Modelling Engine (LIME) model \citep{Brinch2010} that goes beyond the steady-state approximation to include time-variable excitation. The model includes excitation induced by collisions with water molecules and electrons, as well as solar fluorescence pumping rates. The density follows a standard spherically symmetric distribution \citep{Haser1957}. We assumed a gaussian profile for the line, fixed the kinetic temperature $T_{\rm kin}$ = 150\,K, and set a daily water production rates ($Q_{\mathrm{H}_{2}\mathrm{O}}$) at 6.9 $\times$ 10$^{28}$\,mol.\,s$^{-1}$, 7.6 $\times$ 10$^{28}$\,mol.\,s$^{-1}$, and 3.6 $\times$ 10$^{29}$\,mol.\,s$^{-1}$ for April 1, 2 and 3, respectively. The molecular production rates ($Q$), the expansion velocity of the gas $v_{\rm exp}$, and the Doppler shift were left as free parameters and derived using a weighted-$\chi^{2}$ line-fitting analysis. Additional details on the model and its input parameters are provided in \Cref{sec:sublime_param}.

\section{Results}  

    The HCN(3--2) line was identified on each observational day (see \Cref{table:line_int_hcn} and \Cref{fig:hcn_detection_daily}). This is the first HCN detection in a Kreutz sungrazing comet, but also the first molecular detection at millimetre wavelengths in these objects. Other species (CH, CN, C$_2$ and C$_3$) were previously detected in the Kreutz sungrazing comet C/1965 S1 Ikeya-Seki using optical wavelengths \citep{Preston1967,Slaughter1969}.
    
    \begin{figure*}[t!]
    \centering
     \includegraphics[width=1\linewidth]{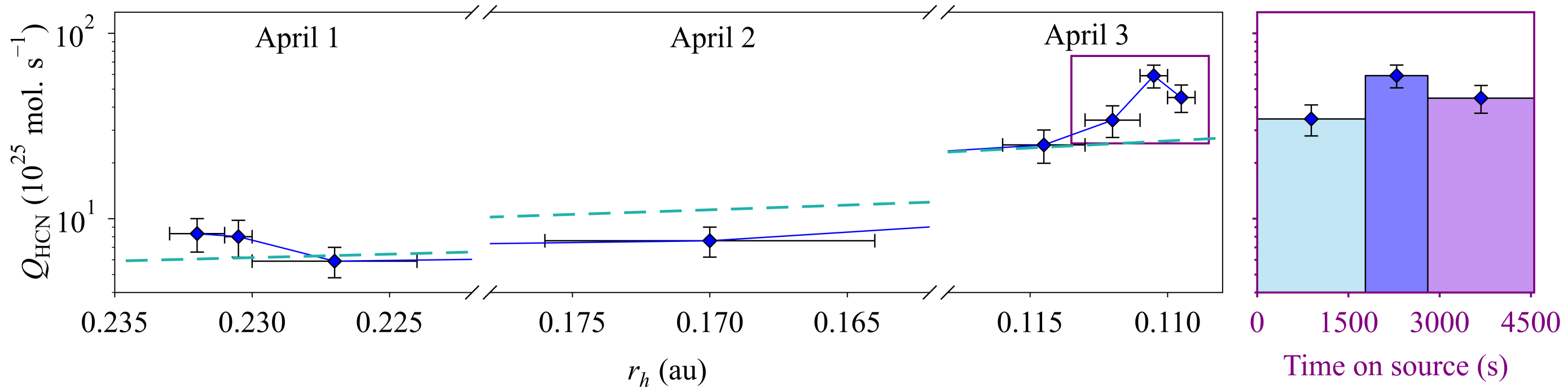} 
    \vspace{-6mm} 
    \caption{Left panel shows the variability of the HCN production rates ($Q_{\mathrm{HCN}}$) with the heliocentric distance ($r_{h}$). The right panel is the time evolution of $Q_{\rm HCN}$ during April 3 outburst (highlighted by the purple rectangle on the left panel). The dotted line shows the best-fit power-law model $Q_{\rm HCN}$ = (0.34 $\pm$ 0.10)$r_h^{(-1.97 \pm 0.14)}$}.
    \label{fig:hcn_vexp_evolution_rh}
    \end{figure*}

    HCN molecular production rates ($Q_{\rm HCN}$) were derived for each observing day by fitting the line profiles with the SUBLIME-1D model described in \Cref{sec:non-LTE}. The resulting molecular spectra are presented in \Cref{fig:hcn_detection_daily} and \Cref{fig:hcn_detection_detail}. The $Q_{\rm HCN}$ values are reported with 1$\sigma$ covariance error bars.  $Q_{\rm HCN}$ increased by 20\% from April 1 to April 2, and by over 350\% from April 2 to April 3. The general increase in $Q_{\rm HCN}$ from one observational day to the next (see \Cref{table:line_int_hcn}) reflects an increasing cometary activity as C/2026 A1 gets closer to the Sun.

    To better understand the variability of the HCN emission, we split the observations from each observational day into subsets (when possible), gathering scans until a confidence level $>4\sigma$ was reached. The results are summarized in \Cref{table:line_int_hcn}, and \Cref{fig:hcn_detection_detail}. The emission varied both across days and within each day. The evolution of $Q_{\rm HCN}$ is shown in the left panel of \Cref{fig:hcn_vexp_evolution_rh}, and clearly presents two outbursts: a weak one on April 1, and a stronger one on April 3. Within the last three measurements on April 3, $Q_{\rm HCN}$ has increased over 75\% for 17\,minutes, followed by a immediate decrease of 25\%. Since the lifetime of HCN at these $r_{h}$ is only $\sim$12\,min, shorter than the integration time (30, 17 and 29\,minutes, see right panel of \Cref{fig:hcn_vexp_evolution_rh}), each spectrum captures predominantly new HCN molecules. Thus, by integrating $Q_{\rm HCN}$ over time (right panel of \Cref{fig:hcn_vexp_evolution_rh}), we probe the total quantity of HCN molecules released during the observations. We find that the comet lost 9.9 $\times$ 10$^{4}$\,kg of HCN during this outburst. If we assume an HCN/H$_2$O production ratio of 0.1\% (the average ratio in comets, \citealt{Biver2024a}), the comet has lost 6.6 $\times$ 10$^{7}$\,kg of H$_2$O. This corresponds to 0.4\% of its total mass, based on a typical nucleus density of 500\,kg\,m$^{-3}$ \citep{Jones2018}. By including the dust dragged along in the process (assuming a dust-to-gas ratio of 2, in the high range of measurements in 67P/Churyumov-Gerasimenko, \citealt{Biver2019}), the total mass loss could reach 1.2\%. If this outburst released a square surface layer of 5\,m thick, this patch would measure 163\,m $\times$ 163\,m for a purely gaseous process, or up to 282\,m $\times$ 282\,m when accounting for the dust. This represents from 7\%, or up to 21\% of the total surface of the comet.

    If we exclude the last three $Q_{\rm HCN}$ measurements in April 3 which are clearly part of an outburst, $Q_{\rm HCN}$ does follow the usual $\propto r_{h}^{-2}$ behaviour (\citealt{Rickman2017}, see \Cref{fig:hcn_vexp_evolution_rh}). The evolution of $v_{\rm exp}$ with $r_{h}$ is discussed in \Cref{sec:vexp}.

    Using estimates of the nucleus size (0.4\,km in diameter,  \citealt{Zhang2026}) and $Q_{\rm H_2O}$, together with a slow-rotator model, we derived daily-averaged active fractions of 175\%, 190\%, 347\%, and 683\% for March 27 (see \Cref{sec:sublime_param} for the associated $Q_{\rm H_2O}$), April 1, 2, and 3, respectively. These values suggest that the activity was sustained by the sublimation of icy grains, with the effective cross-section of the dust coma increasing as $r_h$ decreased, possibly due to dust fragmentation in addition to ongoing nucleus embrittlement and fragmentation.

    We also report the non-detection of HNC, CS, H$_2$CO, CO, OCS and CH$_3$OH (see \Cref{sec:non_detection}). For HNC, the velocity-integrated noise level is 2.7$\sigma$, suggesting a tentative detection.

    \section{Discussion}

    \citet{Sekanina2026} and \citet{Sekanina2026a} support the hypothesis that C/2026 A1 is a second-generation fragment of the original parent, Aristotle's comet (372 BC), and a fragment of the comets AD 363 recorded by Ammianus Marcellinus. Thus, C/2026 A1 would have undergone two perihelion passages within Roche's limit. This HCN detection, the first in a Kreutz sungrazing comet, thus provides direct evidence that cometary volatiles can survive extreme perihelion passages. The high porosity of cometary nuclei results in low thermal inertia, causing solar radiation to sublimate only their surface layers.

    A long-standing question remains regarding the triggering mechanisms of cometary nuclei disruptions. While tidal forces are often invoked, the spontaneous disruption of comets far from Jupiter or the Sun suggests that other mechanisms are at work \citep{Jones2018}. Rapid rotation and torquing forces, potentially driven by irregular outgassing, have been proposed by \citet{Sekanina2015} and \citet{Steckloff2015}. 

    Light curve analysis of this comet during the two months pre-perihelion showed an irregular brightening rate, with an inflection point on February 9--12 going from rapid brightening to a much slower increase \citep{Sekanina2026a}. The absence of a flare-up ruled out fragmentation, leading \citet{Sekanina2026a} to interpret it as an emission source on the nucleus running out of gas. They also attributed orbit determination challenges to erratic outgassing.
    
    Our observations of C/2026 A1 revealed a fast dynamic and highly irregular outgassing, with repeated outbursts (on April 1 and 3). The most significant outburst, on April 3, was captured over $\sim$ 1.5\,hours and caused a 0.4\%--1.2\% total mass loss, potentially triggering the nucleus disintegration observed $< 24$\,h later on April 4. These results support the hypothesis that erratic outgassing, which increases ram pressure, nucleus rotation, and torquing forces, can trigger the disruption of cometary nuclei.

    The study of sungrazing comets depends critically on early discovery, as illustrated by C/2012 S1 (ISON), whose early detection enabled extensive multiwavelength observations. The IRAM 30-m telescope, with its ability to observe at solar elongations as small as 5°, is particularly suited for investigating cometary chemistry near perihelion, when ALMA and SOHO/SWAN face more restrictive Sun-avoidance constraints.

\begin{acknowledgements}
     We thank Dr. Karl Friedrich Schuster, IRAM director, for granting a DDT proposal, as well as Dr. Pablo Torne, Gloria Liñán Rodríguez, and Dr. Brisa Mancillas for support in this project success, and all IRAM 30-m staff. IRAM is supported by the Institut Nationnal des Sciences de l’Univers (INSU) of the French Centre national de la recherche scientifique (CNRS), the Max-Planck-Gesellschaft (MPG, Germany) and the Spanish Instituto geográfico nacional (IGN). A.G.P. and V.M.R. acknowledge support from the grant CNS2023-144464 funded by MICIU/AEI/10.13039/501100011033 and by “European Union NextGenerationEU/PRTR”, and from the grant PID2022-136814NB-I00 by the Spanish Ministry of Science, Innovation and Universities/State Agency of Research MICIU/AEI/10.13039/501100011033 and by ERDF, UE. M. C. acknowledges financial support from the NSF grant AST-2509096, and NASA’s Planetary Science Division Internal Scientist Funding Program through the Fundamental Laboratory Research work package (FLaRe).

\end{acknowledgements}

   \bibliographystyle{aa} 
   \bibliography{bib_test} 

\begin{appendix}
	\nolinenumbers

\section{Observation conditions}\label{sec:obs_conditions}

The observing window was set from 09:00 to 16:30 UTC, during favourable elevation of the comet ($>30\degree$). Observations after April 3 were not feasible because the solar elongation angle was $<5\degree$, the safety limit of the IRAM 30-m telescope. Due to windy conditions, observations started only at 11:00 UTC on April 2 and 10:00 UTC on April 3. Calibration issues on April 2 caused by atmospheric instability prevented observations between 13:45 and 16:00 UTC. A summary of observational parameters are provided in \Cref{table:observational_parameters}.

{\color{black}We attempted to determine the position of the comet through the HCN peak emission in a coarse map, and estimated it to be within $\pm$7.5\arcsec in right ascension. However, for a reasonable integration time, the signal was not bright enough to determine the best pointing correction, so we decided to keep the position given by the JPL ephemeris.}

    \begin{table*}[t]
    \caption{Observational parameters of the C/2026 A1 (MAPS) comet, with the heliocentric distance ($r_h$), geocentric distance ($\Delta$), elongation angle (Sun--observer--target), phase angle (Sun--target--observer), and precipitable water vapour (pwv).}
    \label{table:observational_parameters}
    \centering
    \begin{tabular}{c c c c c c c c}
    \hline\hline
    UTC date & $< r_{h}>$ &  $< \Delta>$ &  $<$ Elong. $>$ &  $<$ Phase $>$ & pwv & Integration time & Frequency range \\   
    (yyyy/mm/dd.d-dd.d) & (au) & (au) &  ($\degree$) & ($\degree$) & (mm) & (min) & (GHz) \\    
    \hline                  
     2026/04/01.41 - 01.58 & 0.229  & 1.090  &  11.56  & 61.1   &   1.6--2.7 & 288 & 248.3--256.4 ; 264.3--272.4 \\ 
     2026/04/01.61 - 01.69 & 0.221  & 1.087  &  11.15  & 61.0   &   2.1--3.2 & 138 & 224.6--232.7 ; 240.6--248.7\\ 
     \hline
     2026/04/02.49 - 02.57 & 0.173  & 1.074  &   8.67  & 60.4   &   1.0--2.4 & 132 & 248.3--256.4 ; 264.3--272.4 \\ 
     2026/04/02.67 - 02.69 & 0.164  & 1.071  &   8.21  & 60.2   &   0.9--1.0 & 48 & 248.3--256.4 ; 264.3--272.4 \\ 
     \hline
     2026/04/03.45 - 03.54 & 0.112  & 1.055  &   5.44  & 57.7   &   1.6--2.1 & 162 & 248.3--256.4 ; 264.3--272.4 \\ 
     2026/04/03.59 - 03.61 & 0.105  & 1.053  &   5.06  & 57.2   &   2.0--2.2 & 49 & 224.6--232.7 ; 240.6--248.7\\ 
    \hline
    \hline
    \end{tabular}
    \end{table*}

   \section{SUBLIME-1D parameters} \label{sec:sublime_param}

       \begin{table}[t!]
       \renewcommand{\arraystretch}{1.3}
        \caption{Photodissociation rates at 1\,au ($\beta_0$) employed in the SUBLIME-1D model.}
        \label{table:sublime_parameters}
        \centering                  
        \begin{tabular}{c c c} 
        \hline\hline          
        Name &   $\beta_0$    &   Reference \\         
	         			 &  (10$^{-5}$\,s$^{-1}$)    &  \\
        \hline                     
            HCN     & 1.7                           &   \citet{Hrodmarsson2023}\\    
            \hline
            HNC     & 1.7                           &   Assumed from HCN measurements\\
            \hline
            \multirow{2}{*}{CS}      & 0.31                          &   \citet{Hrodmarsson2023}\\
                    & 2.5                           &   \citet{Biver2011}\\
            \hline
            \multirow{2}{*}{H$_2$CO} & 0.47                          &   \citet{Hrodmarsson2023}\\
                    & 20                            &   \citet{Bockelee-Morvan1992}\\
            \hline
            CO      & 0.19                          &   \citet{Hrodmarsson2023}\\
            \hline
            OCS     & 0.35                          &   \citet{Hrodmarsson2023}  \\
            \hline
            \multirow{2}{*}{CH$_3$OH} & 0.69                         &   \citet{Hrodmarsson2023} \\
            												& 1.2                         &   \citet{Huebner1992} \\
        \hline\hline                           
        \end{tabular}
    \end{table}

   Input parameters for the determination of molecular production rates using the SUBLIME-1D model include the kinetic temperature of the gas, water production rate (assumed to be the most abundant species in the coma), and the electronic fraction \citep{Cordiner2022}.

   \subsection{Kinetic temperature within the coma}

    In comets observed at similar $r_{h}$ (C/2002 V1, C/2002 X5 and C/2006 P1), the kinetic temperature was varying between 80\,K and 220\,K \citep{Biver2011}. In the absence of methanol lines to constrain the gas kinetic temperature, we decided to fix the kinetic temperature at 150\,K. A variation of $\pm$ 50\,K was affecting $Q_{\rm HCN}$ by $< 10\%$ on April 1 and 2, and $< 30\%$ on April 3.

    \subsection{Water production rates}

    {\color{black}SOHO/SWAN observations of C/2026 A1 on March 27, 2026 resulted in an estimate of $Q_{\rm H_2O}$ = 2 $\times$ 10$^{28}$ \,mol\,s$^{-1}$ (M. R. Combi, private communication). We also unsuccessfully searched for OH with the Nançay radio telescope from 15 to 30 March. As there are no measurements of $Q_{\rm H_2O}$ for the observation date (April 1--3)}, we have estimated the $Q_{\rm H_2O}$ assuming an HCN abundance relative to water of 0.1\% on each date, the mean $Q_{\rm HCN}/Q_{\rm H_{2}O}$ in comets \citep{Biver2024a}. We obtain $Q_{\rm H_{2}O}=6.9\times 10^{28}$\,mol.\,s$^{-1}$ on April 1, $Q_{\rm H_{2}O}=7.6\times 10^{28}$\,mol.\,s$^{-1}$ on April 2, and $Q_{\rm H_{2}O}=3.6\times 10^{29}$\,mol.\,s$^{-1}$ on April 3, and $Q_{\rm H_{2}O}=1.1\times 10^{29}$\,mol.\,s$^{-1}$ by averaging HCN observations on the three days. The latter value was used to derive upper limits on the molecular production rates for the other species. {\color{black}These values, together with the $Q_{\rm H_2O}$ measured by SOHO/SWAN on March 27, are consistent with increasing cometary activity as the heliocentric distance decreases.} A variation of $Q_{\rm H_2O}$ by $\pm$ 50\% impacts $Q_{\rm HCN}$ by $\pm$ 10\%, which falls into the $1\sigma$ covariance error bars. Thus, we consider that the assumed $Q_{\rm H_2O}$ will not impact the conclusions in this paper.
    
    These values are relativity small compared with those derived by \citet{Biver2011}, with $Q_{\rm H_{2}O}=3 \times 10^{30}$\,mol.\,s$^{-1}$ for C/2002 X5 at $r_{h}=0.214$\, au, $Q_{\rm H_{2}O} \sim 2\times 10^{30}$\,mol.\,s$^{-1}$ for C/2002 V1 at $r_{h} \sim 0.1$\, au ; and $Q_{\rm H_{2}O}= [5 - 40] \times 10^{30}$\,mol.\,s$^{-1}$ for C/2006 P1 at $r_{h} = [0.21 - 0.25]$\, au. 

    \subsection{Electronic fraction}
    
    We fixed the electronic fraction at 0.5, and its variation by $\pm$ 50\% affects $Q_{\rm HCN}$ by 2--12\%. The derived $Q_{\rm HCN}$ are less sensitive to the electronic fraction at smaller $r_{h}$ because a larger fraction of molecules photodissociate before leaving the collisional region, where the population of energy levels are thermalised at the kinetic temperature of the gas (see \Cref{sec:non-LTE}).

    \subsection{Photodissociation rate}
    
    At these very short heliocentric distances, the determination of the molecular production rates is very sensitive to the assumed photodissociation rates $\beta$. The photodissociation rates at 1\,au $\beta_0$ employed in the model for each molecule can be found in \Cref{table:sublime_parameters}. In the Leiden database, the measured photodissociation Lyman-$\alpha$ cross-section, from which $\beta_{\rm HCN}$ obtained, is given to be accurate within 30\%. Within this $r_{h}$ range, the $Q_{\rm HCN}$ are affected by 25--30\% on all dates. 

    \subsection{Molecular data}
    
    The molecular data employed in the modelling are mainly available in databases. The collisional data for collisions with H$_2$O were used when available (HCN\footnote{This dataset was also used to model HNC.} \citealt{Zoltowski2025}, CS \citealt{GodardPalluet2026}, CO \citealt{Faure2020}), with H$_2$ if not (H$_2$CO \citealt{Wiesenfeld2013}), and they were calculated with global cross-sections approximation if none of the previously mentioned data were available (OCS, CH$_3$OH\footnote{The CH$_3$OH--H$_2$ collisional data provided by \citet{Dagdigian2023} could not be used because they did not cover enough energy levels.} with $\sigma = 5\times10^{-14}$\,cm$^{-2}$ \citealt{Crovisier1987,Bockelee-Morvan1987}). For the collisions with electrons, the data were taken from the literature (HCN and HNC \citealt{Faure2007}, CS \citealt{Varambhia2010}, H$_2$CO \citealt{Gerin2024}, CH$_3$OH \citealt{Guzman2013}), or calculated with the Born dipole approximation (CO, OCS) following the work of \citet{Itikawa1972}. The datasets were generally taken from the Excitation of Molecules and Atoms for Astrophysics database (EMAA, \citealt{Faure2025}). 
    
    The solar pumping rates were taken from the Planetary Spectrum Generator\footnote{\url{https://psg.gsfc.nasa.gov/index.php}}, and the photodissociation rates $\beta$ were obtained from the Lyman-$\alpha$ cross-sections provided in the Leiden database \citep{Hrodmarsson2023}\footnote{\url{https://home.strw.leidenuniv.nl/~ewine/photo/}}, multiplied by the photon flux at the dates of the observations from the Laboratory for Atmospheric and Space Physics.\footnote{\url{https://lasp.colorado.edu/data/timed_see/composite_lya/version3/composite_lya.dat}} For the parent length $L_p$ (see \Cref{sec:non_detection}), which describes the distance from the nucleus at which the daughter species are being formed, we assumed HCN, HNC, CO, OCS and CH$_3$OH to be parent species ($L_p=0$\,km), CS to be produced from CS$_2$ at a rate of $1.9\times10^{-3}$\, s$^{-1}$ at 1\,au \citep{Heays2017} and H$_2$CO from a parent (unidentified) species at a rate $k = 7\times10^{-4}$\, s$^{-1}$ at 1\,au, following the work of \citet{Cordiner2023}.

\section{Evolution of the expansion velocity}\label{sec:vexp}

\begin{figure*}[t!]
    \centering
     \includegraphics[width=1\linewidth]{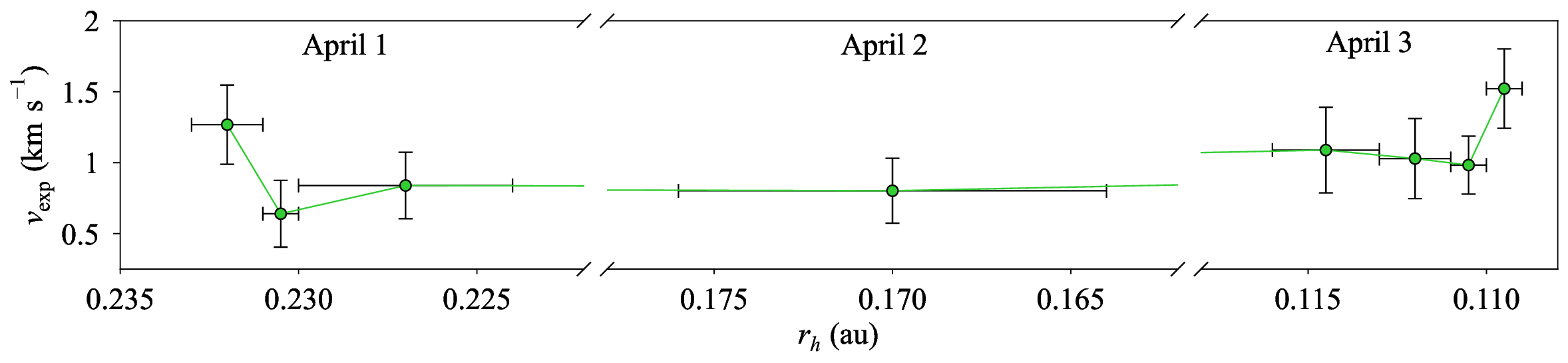} 
        \vspace{-6mm} 
    \caption{Variation of the expansion velocity ($v_{\mathrm{exp}}$, green circles) with the heliocentric distance ($r_{h}$).}
    \label{fig:vexp_rh}
    \end{figure*}

    As the comet approached the Sun, the expansion velocity increased (see \Cref{fig:vexp_rh}). However, during the outburst, the expansion velocity of the gas first decreases, reaching minimum values near the HCN emission peaks, and then increases again. This unusual behaviour could be explained by geometric considerations. A collimated jet directed toward the Sun would reduce the observed line width, leading to an underestimation of the gas expansion velocity if we assume isotropic outgassing. However, due to the large uncertainties in both $v_{\rm exp}$ and the Doppler shift, it is difficult to analyse these effects further.

\section{Upper limits for other species}\label{sec:non_detection}

    \subsection{Non-detections of HNC, CS, CO, H$_2$CO, OCS and CH$_3$OH}

\begin{figure*}[t]
    \centering
    \begin{minipage}{\linewidth}
    \captionof{table}{Physical parameters derived for HNC, CS, H$_2$CO, CO, OCS and CH$_3$OH.}
    \label{table:non_detection}
    \centering 
        \renewcommand{\arraystretch}{1.3}
    \begin{tabular}{c c c c c c c c}
    \hline\hline 
    
    Molecule & Transition & Frequency & $\int T_{\rm mb}d\nu$ &  S/N      & $Q$  & $Q$/$Q_{\rm HCN}$ & $Q$/$Q_{\rm HCN}$ in comets\\   
             &            & (GHz)     & (mK\,km\,s$^{-1}$) &    & ($10^{25}$ mol.\,s$^{-1}$)  &  &  \\    
    \hline 
    \multirow{2}{*}{HNC}     & \multirow{2}{*}{3--2}                 & \multirow{2}{*}{271.9811} & 39.6 $\pm$ 15$^{(a)}$  & 2.7 & 2.8 $\pm$ 1.0 & 0.25 & \multirow{2}{*}{0.01--0.45}\\
            &                      &          & $<$ 40  & $<$ 3 & $<$ 3.2  & $<$ 0.29 &              \\ 
    \hline
    \multirow{2}{*}{CS}      & \multirow{2}{*}{5--4}                 & \multirow{2}{*}{244.9356} & \multirow{2}{*}{$<$ 65}  & \multirow{2}{*}{$<$ 3} & $<$ 6.2$^{(b)}$ & $<$ 0.6 & \multirow{2}{*}{0.2--3.2}      \\
            &                      &         &         &       & $<$ 24.2$^{(c)}$ & $<$ 2.2 &            \\
    \hline
    \multirow{2}{*}{H$_2$CO} & \multirow{2}{*}{$3_{12}$--$2_{11}$}   & \multirow{2}{*}{225.6978} & \multirow{2}{*}{$<$ 53}  & \multirow{2}{*}{$<$ 3} & $<$ 19.3$^{(b)}$  & $<$ 1.7 & \multirow{2}{*}{1.3--13}    \\
            &                      &         &         &       & $<$ 388$^{(d)}$  & $<$ 122 &           \\
    \hline
    CO      &  2--1                & 230.5380 & $<$ 56  & $<$ 3 & $<$ 2450 & $<$ 227 & 8--270         \\ 
    \hline
    OCS     & 20--19               & 243.2180 & $<$ 71  & $<$ 3 & $<$ 112 & $<$ 10.2 & 0.43--3.5      \\ 
    \hline
     \multirow{2}{*}{CH$_3$OH}	&	\multirow{2}{*}{5$_{K}$--4$_{K}$, $K$ = 0, 1, 2, 3}	&	 \multirow{2}{*}{241.7--243.9}		&	\multirow{2}{*}{$<$ 188}		& \multirow{2}{*}{$<$ 3}	&	$<$ 106$^{(b)}$	&	$<$ 9.6	& 	\multirow{2}{*}{4--64}	\\
   	&  &	& 	&  		& 	$<$ 160	$^{(d)}$&	$<$ 15	& 		\\

    \hline
    \hline
    \end{tabular}
    \begin{center}
    \end{center}\vspace{-0.7cm}
    {\small
    \begin{flushleft}
    $^{(a)}$ This line area was calculated with $v_{\rm exp}$ = 1.13 $\pm$ 0.40\,km\,s$^{-1}$, obtained by fitting the HNC line profile using the SUBLIME model. Upper limits are calculated using $v_{\rm exp}$ obtained from the HCN line profile, as described in the text.\\
    $^{(b)}$ Using $\beta_{0}$ from \citet{Hrodmarsson2023}.\\
    $^{(c)}$ Using $\beta_{0}$ from \citet{Bockelee-Morvan1992}.\\
    $^{(d)}$ Using $\beta_{0}$ from \citet{Huebner1992}.
    \end{flushleft}
    }
    \end{minipage}

    \vspace{0.5cm}
    
    \includegraphics[width=\linewidth]{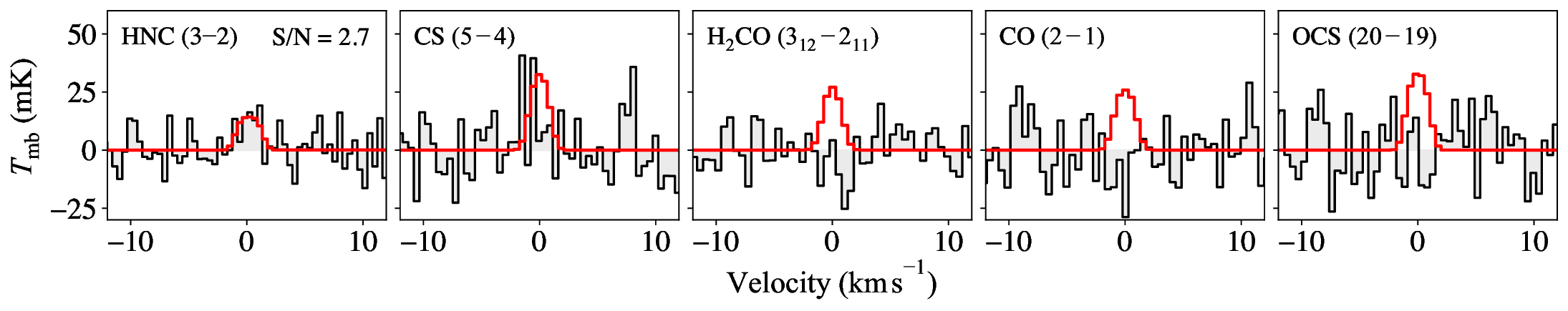} 
    \vspace{-6mm} 
    \captionof{figure}{Observed (grey lines) and modelled (red lines) main beam temperature ($T_{\rm mb}$) in the comet velocity rest-frame for the HNC(3--2), CS(5--4), H$_2$CO($3_{12}$--$2_{11}$), CO(2--1), and OCS(20--19) non-detected emission lines.}
    \label{fig:non_detection} 
\end{figure*}

\begin{figure}[t!]
	\centering
	\includegraphics[width=1\linewidth]{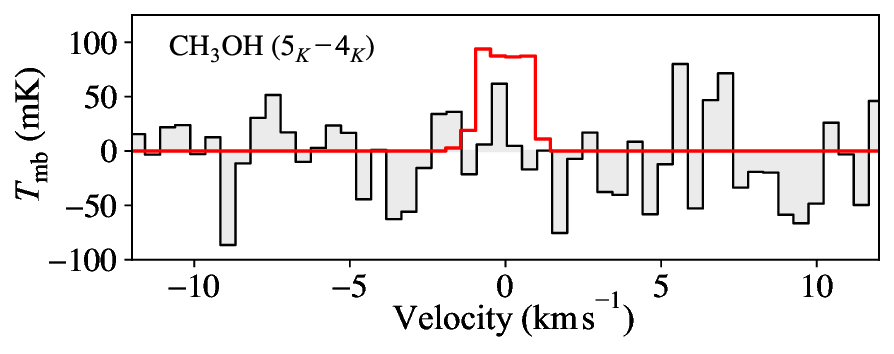} 
	\vspace{-6mm} 
	\caption{Observed (grey lines) and modelled (red lines) main beam temperature ($T_{\rm mb}$) in the comet velocity rest-frame obtained by stacking 11 emission lines CH$_3$OH (5$_K$--4$_K$) between 241.7 and 243.9\, GHz. The model corresponds to a line area with S/N = 3, derived assuming $Q_{\rm CH_3OH}$ = 1.06 $\times$ 10$^{27}$\,mol.\,s$^{-1}$.}
	\label{fig:stacking_ch3oh}
\end{figure}

     The production rates for HNC, CS, CO, H$_2$CO and OCS were derived by fitting a line whose integrated area corresponds to a 3$\sigma$ noise level (see \Cref{fig:non_detection}). For CH$_3$OH, we derived an upper limit by stacking 11 CH$_3$OH(5$_{K}$--4$_{K}$) lines between 241.1\,GHz and 243.9\,GHz, corresponding to $K$ = 0, 1, 2, 3. The modelled line emission shown in \Cref{fig:stacking_ch3oh} was obtained from the same approach, with an integrated area corresponding to a 3$\sigma$ noise level. The excitation models were run by fixing $v_{\rm exp}$ = 0.94\,\kms (derived from the mean HCN line profile, calculated by averaging all available HCN observational data), and $Q_{\rm H_{2}O}=1.1\times 10^{29}$\,mol.\,s$^{-1}$ (see \Cref{sec:sublime_param}). 

    The upper limits for the $Q$/$Q_{\rm HCN}$ ratios of all species, exhibited in \Cref{table:non_detection}, generally fall within the ranges observed for other comets \citep{Biver2026}. However, CS and H$_2$CO lie at the lower end of these ranges, which is inconsistent with their expected formation mechanisms. For CS, the $Q_{\rm CS}$/$Q_{\rm HCN}$ ratio in comets follows a $r_{h}^{-1}$ variation \citep{Biver2011}, suggesting a temperature-dependent production mechanism in addition to CS$_2$ photolysis. As for H$_2$CO, it is expected to form from the thermal degradation of organic matter, such as polyoxymethylene \citep{Fray2006}. Thus, both species should be more abundant at smaller $r_{h}$. 
    
    A potential origin for this effect could be the photodissociation rates $\beta_{0}$ employed in the different works. In the case of H$_2$CO, quite a few values have been reported in the literature. \citet{Mentall1971} measured a UV photodissociation cross-section, leading to $\beta_0$(H$_2$CO) = $1.5 \times 10^{-5}$\, s$^{-1}$. \citet{Huebner1979} measured the photodissociation rates for 5 branches, and found a total photodissociation rate $\beta_0$(H$_2$CO) = $2.8 \times 10^{-4}$\, s$^{-1}$. \citet{Bockelee-Morvan1992} adopted a $\beta_0$(H$_2$CO) = $2 \times 10^{-4}$\, s$^{-1}$, accounting for the fact that the dissociation of H$_2$CO is sensitive to its rotational distribution, which differs between the coma, and room temperatures at which the measurements were realised. The latter value is close to $\beta_0$(H$_2$CO) = $1.95 \times 10^{-4}$\, s$^{-1}$ of \citet{Boice1991}, who obtained their value through multi-fluid modelling of 22P/Kopff comet. The value of \citet{Bockelee-Morvan1992} was commonly used in future studies. Recently, \citet{Hrodmarsson2023} provided a new photodissociation cross-section measurement, from which we obtained our $\beta_0$(H$_2$CO) = $4.7 \times 10^{-6}$\, s$^{-1}$. This $\beta_0$ is 40 times lower than the values used in previous studies. In their measurement, only the principal photodissociation branch is accounted for (H$_2$CO + $\nu$ $\rightarrow$ CO + H$_2$). However, two other branches can also significantly contribute (producing HCO + H, and CO + 2H, \citealt{Huebner1979}). It is thus possible that this value underestimates the efficiency at which H$_2$CO is destroyed by solar radiation. Using the value of \citet{Bockelee-Morvan1992}, $Q_{\rm H_2CO}$ has increased by a factor 20, and now lies within the usual ranges of H$_2$CO abundances in comets (see \Cref{table:non_detection}).

    In the case of CS, its photodissociation cross-section has been estimated at $5.36 \times 10^{-17}$\, cm$^{2}$ by \citet{vanDishoeck1988} combining theoretical and experimental work of \citet{Bruna1975} and \citet{Stark1987}, respectively. Based on the recent work of \citet{Xu2019}, it has be refined by \citet{Hrodmarsson2023} to 6.46 $\times$ 10$^{-18}$\, cm$^{2}$, from which we obtained the $\beta_0$(CS) employed in this work. Several $\beta_0$(CS) values were estimated from observational data, and they were varying from 10$^{-5}$--10$^{-4}$\,s$^{-1}$, with a overall agreement for a value around $(1-3) \times 10^{-5}$\,s$^{-1}$ (see \citealt{Boissier2007} and \citealt{Biver2011} for further details). By using $\beta_0$(CS) = $2.5 \times 10^{-5}$\,s$^{-1}$ from \citet{Biver2011}, our $Q_{\rm CS}$ has increased by almost a factor 4, and now also relies within the usual CS abundance ranges (see \Cref{table:non_detection}).
    
    We also tested the sensitivity of $Q_{\rm CH_3OH}$/$Q_{\rm HCN}$ to $\beta_0$, as the values reported in previous work can be twice as high \citep{Huebner1992}. As for CS and H$_2$CO, $Q_{\rm CH_3OH}$ is sensitive to $\beta_0$, and has increased by 51\% when using $\beta_0$ = $1.2 \times 10^{-5}$\,s$^{-1}$ from \citet{Huebner1992}, as exhibited in \Cref{table:non_detection}.
    
    In the work of \citet{Biver2011}, the $Q_{\rm CH_3OH}$/$Q_{\rm HCN}$ has been found to be $\propto r_{h}^{0.5}$ at low $r_{h}$, a trend already observed in the study of C/1996 B2 (Hyakutake) \citep{Biver1999}. They also found this ratio to be relatively sensitive to the assumed temperature of the gas, with a variation by up to a factor of 2 in C/2006 P1. In our case, a variation of $T_{\rm kin}$ by $\pm$ 50\,K led to a variation in $Q_{\rm CH_3OH}$ by $\pm$ 26\%. Therefore, the assumption we made on the kinetic temperature of the gas is expected to have a relatively weak impact on this result.

    \subsection{The HNC/HCN abundance ratio}\label{sec:HNC/HCN}
    
\begin{figure*}
    \centering
        \includegraphics[width=\textwidth]{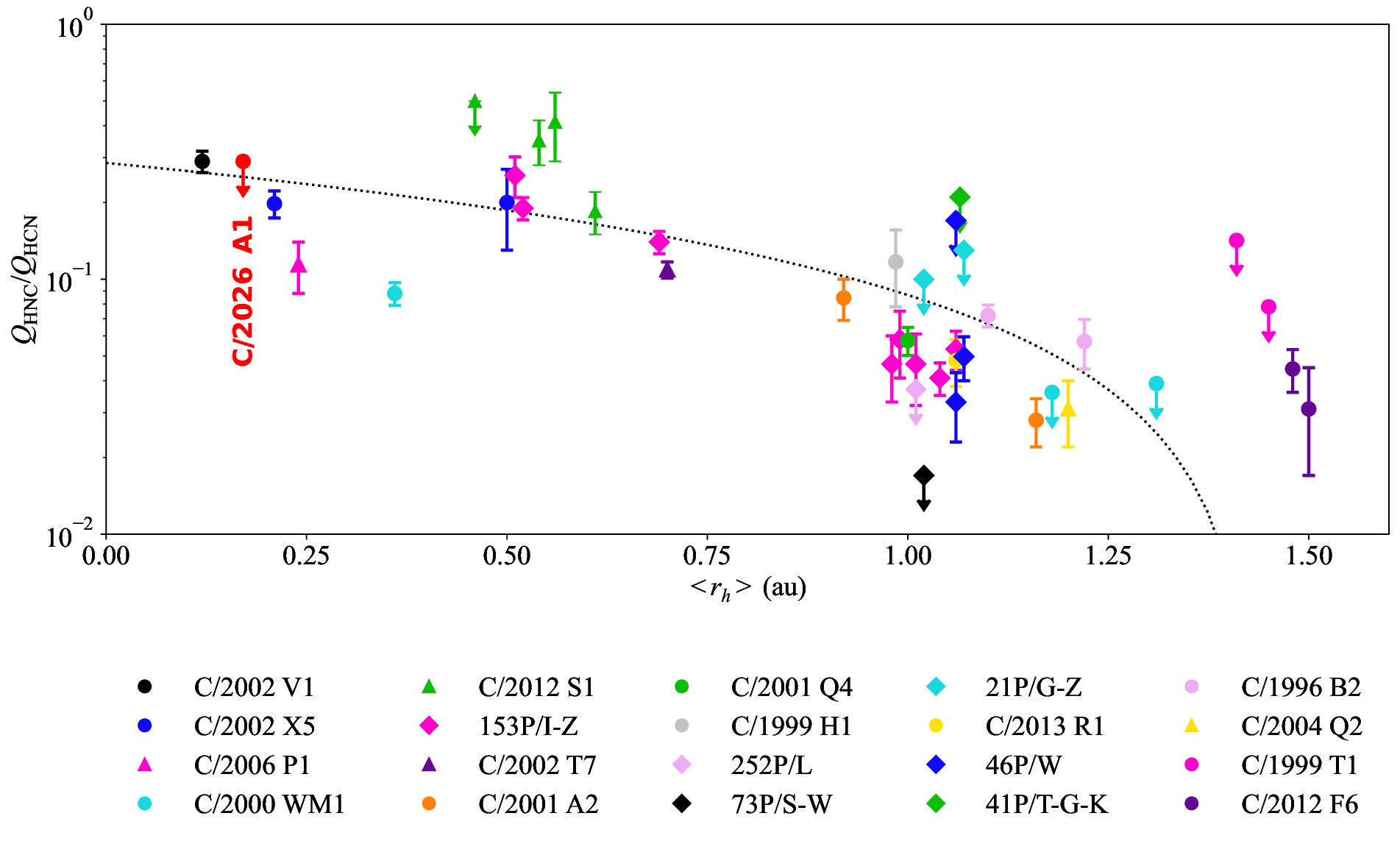} 
    \caption{Evolution of the $Q_{\rm HNC}/Q_{\rm HCN}$ ratio with the heliocentric distance $r_{h}$. The comets are sorted by type: Jupiter comets (diamonds), dynamically new (triangles) and dynamically old (circles) Oort cloud comets. The data points were added from \citet{Lis2008, Biver2011,DeVal-Borro2012,DeVal-Borro2013,Agundez2014a,Cordiner2014,DelloRusso2016,Bogelund2017a,Coulson2017,Coulson2020,Biver2021} and this work.
    }
    \label{fig:hcn_hnc_ratio}
    \end{figure*}

    HNC is thought to originate from organic refractory material \citep{Lis2008,Cordiner2017}, released from confined regions on the nucleus. An evidence of this formation path is the evolution of the $Q_{\rm HNC}/Q_{\rm HCN}$ ratio with $r_{h}$ \citep{Lis2008}. In the literature, 40 measurements are now available for 20 comets. These measurements are reported in \Cref{fig:hcn_hnc_ratio}, including the new measurement in C/2026 A1 reported in this work.
    
    A linear law was suggested by \citet{Lis2008} to represents the behaviour of $Q_{\rm HNC}/Q_{\rm HCN}$ with $r_{h}$. By including all new measurements reported in \Cref{fig:hcn_hnc_ratio} (excluding upper limits), we obtain a best fit given by $Q_{\rm HNC}/Q_{\rm HCN}$ = ($-0.20\pm0.04$)$r_{h}$ + (0.29$\pm$0.03), with a reduced $\chi^{2}$ = 16.6. We also tested a power law relation, and obtain $Q_{\rm HNC}/Q_{\rm HCN}$ = (0.10$\pm$0.20)$r_{h}^{(-0.56\pm1.73)}$, and a reduced $\chi^{2}$ = 19.7. With the available measurements, the linear model provides a better description of the data, as indicated by its lower reduced $\chi^{2}$, and the poor constraint on the exponent of the power law fit. Nevertheless, both relations indicate an increasing $Q_{\rm HNC}/Q_{\rm HCN}$ ratio with decreasing $r_h$. This behaviour is consistent with the hypothesis that HNC is increasingly efficiently produced as the heliocentric distance decreases, i.e. as the solar radiation field increases and degradation mechanisms become more efficient. 

\end{appendix}
\end{document}